\documentclass[aps, prd, twocolumn, superscriptaddress, nofootinbib, showpacs]{revtex4}
\usepackage{graphicx}% Include figure files
\usepackage{dcolumn}% Align table columns on decimal point
\usepackage{amssymb}% outlined letters
\usepackage{amsmath}% outlined letters
\usepackage{array}
\usepackage{float}

\begin{document}

%Macros
\newcommand{\Eq}[1]{\mbox{Eq. (\ref{eqn:#1})}}
\newcommand{\Fig}[1]{\mbox{Fig. \ref{fig:#1}}}
\newcommand{\Sec}[1]{\mbox{Sec. \ref{sec:#1}}}

\newcommand{\PHI}{\phi}
\newcommand{\PhiN}{\Phi^{\mathrm{N}}}
\newcommand{\vect}[1]{\mathbf{#1}}
\newcommand{\Del}{\nabla}
\newcommand{\unit}[1]{\;\mathrm{#1}}
\newcommand{\x}{\vect{x}}
\newcommand{\ScS}{\scriptstyle}
\newcommand{\ScScS}{\scriptscriptstyle}
\newcommand{\xplus}[1]{\vect{x}\!\ScScS{+}\!\ScS\vect{#1}}
\newcommand{\xminus}[1]{\vect{x}\!\ScScS{-}\!\ScS\vect{#1}}
\newcommand{\diff}{\mathrm{d}}

\newcommand{\be}{\begin{equation}}
\newcommand{\ee}{\end{equation}}
\newcommand{\bea}{\begin{eqnarray}}
\newcommand{\eea}{\end{eqnarray}}
\newcommand{\vu}{{\mathbf u}}
\newcommand{\ve}{{\mathbf e}}

        \newcommand{\vU}{{\mathbf U}}
        \newcommand{\vN}{{\mathbf N}}
        \newcommand{\vB}{{\mathbf B}}
        \newcommand{\vF}{{\mathbf F}}
        \newcommand{\vD}{{\mathbf D}}
        \newcommand{\vg}{{\mathbf g}}
        \newcommand{\va}{{\mathbf a}}

%=====================================================================
%=====================================================================
%=====================================================================
\title{Differentiating Between Modified Gravity Theories in the Solar System}
%Testing Different Formulations of Modified Newtonian Dynamics in the Solar System}
\author{Ali Mozaffari}
\affiliation{
Theoretical Physics Group, Imperial College, London, SW7 2BZ}

\email{ali.mozaffari05@imperial.ac.uk}

\date{\today}
%\baselineskip=14truept
\begin{abstract}
{Building on previous work, we re-examine the possibility of testing MOdified Newtonian Dynamics near the saddle points of gravitational potentials in the Solar System, through an extension of the forthcoming LISA Pathfinder mission.  We extend present analysis to include quasi-linear formulations of these theories, resulting from fully relativistic modified gravity theories.  Using similar quantitative and qualitative tools, we demonstrate that in general, both the instrumental response and typical Signal to Noise Ratios for such a test will be different.  Finally we investigate constraints from a negative result and parameterised free functions.}
\end{abstract}
\maketitle
%\tightenlines

\bigskip

\subsubsection{Introduction} The standard model of cosmology is composed
of General Relativity and $\Lambda$CDM which has had great
success in matching large scale observations in astronomy and cosmology.
 However attempts to date to directly detect candidate dark matter particles have been have been at best unclear and at worst fruitless.  Over the past few decades, attempts to reconcile anomalous observations of galactic dynamics without resorting to dark matter have been the driving forces behind modified gravity theories. These have centred around a prescription for a modified force law~\cite{Milgrom:1983ca}, which reduce to Newtonian dynamics at ``large'' accelerations and modified dynamics at ``small'' accelerations (large and small here are relative to the Milgrom characteristic acceleration taken
here as $a_0 \approx 10^{-10}$ ms$^{-2}$). In doing so, the observed flattening of galaxy rotation curves and the empirical Tully-Fisher relation can both be satisfied~\cite{MONDreview}.  Recent work on an experimental observation of theories with a preferred acceleration scale~\cite{bekmag,bevis, ali} suggest that anomalously large tidal stresses should be present around the gravitational saddle points (SP) scattered
throughout the Solar System.  The proposed LISA Pathfinder (LPF) mission~\cite{LISA}
could be capable of serving as an accelerometer of unprecedented accuracy.
 The instrument, essentially, consists of two test masses, orientated such that they are
 in a free fall, with a laser interferometer between them.  In this way, it is possible to measure the differential acceleration between the two bodies and hence infer the tidal stresses (by factoring the inter mass spacing).  With such an instrument, we suggest it could be possible to observe the onset of MONDian behaviour (rather than at the scales present at galaxies, where MONDian behaviour will dominate the dynamics) in the low acceleration regime around these SP.  The purpose of this paper is extend previous analyses done for the modified gravity saddle point science case, based upon a scenario where an LPF mission extension is granted. This would involve redirecting the spacecraft from its location parked at Lagrange point L1 to a saddle of the  Sun-Earth-Moon system~\cite{companion} once its nominal mission is completed.  Previous analysis have considered non-linear theories with modified
Poisson equations, which naturally give rise to modified gravity behaviour
(as seen most clearly in~\cite{aliscaling}).  A review of the different
non-relativistic limits of preferred acceleration scale theories~\cite{ali}
categorised these as types I, II and III (we summarise these briefly in Appendix \ref{theory}).  In this work, we examine type II theories, where the dynamics are governed by a physical potential $\Phi$, composed of $\Phi = \Phi_N + \phi$, with the usual \be \Del^2 \Phi_N = 4\pi G \rho\ee  The field $\phi$ is ruled by a driven linear Poisson equation \be \Del^2 \phi = \frac{k}{4\pi} \Del \cdot (\nu(w)\Del\Phi_N) \label{type II Laplacian}\ee where the argument of free function $\nu$ is given by \be w = \left(\frac{k}{4\pi}\right)^2\frac{|\Del\Phi_N|}{a_0} \ee and we require that $\nu \sim 1/\sqrt{w}$ for $w \ll 1$ and that $\nu \rightarrow$ constant when $w \gg 1$.  Consider now just writing \be \Del^2 \Phi = \Del\cdot(\hat{\nu}(w)\Del\Phi_N)\ee where $\hat{\nu} = 1 + \frac{\kappa}{4\pi}\nu$.
 We can divide our theories into two distinct subclasses, where crucially
 \bea \nu \rightarrow 0 && \text{IIA}\\ \nu \rightarrow 1 && \text{IIB}\eea  Type IIB theories in the large acceleration regime have $G_{ren} = G(1 + \kappa/4\pi)$, due to the scalar field mimicking a rescaled Newtonian
potential $\phi \rightarrow \frac{\kappa}{4\pi}\Phi_N$.  In IIA theories, it can be argued we need only a single physical potential $\Phi$ and Newtonian field $\Phi_N$ plays only an auxiliary role (and so
$\kappa$ need not appear at all) - meaning there is no $G$ renormalisation.
 As pointed out in~\cite{ali}, the effect of which is that the triggering of MONDian effects would happen at $a_{trig} = a_0$ (at a distance of $r \sim 2.2$m from the saddle).  With such a tiny bubble, these theories would escape the net of an LPF test, at least for the Earth-Sun system.  With this in mind, we will stick to type IIB theories here, the case for testing other types of MONDian theory
have been considered separately~\cite{ali,typeIIpaper}.  Similarly, we stress that it is important that we consider formulations wedded to fully relativistic theories (rather than just a scalar Lagrangian theory like AQUAL~\cite{aqual}), even though we are not directly testing these full theories themselves. Their cosmologies however, do provide the necessary constraints on the gravitational constant $G$, the {\it bare} value of $G$ appears in the Friedman equations and so from BBN constraints can be fixed.  As we will see, this issue is not a trivial one in modified gravity theories.  Deriving the full field equations and taking the weak field limit as necessary allows us to find the modified Poisson equations for these theories consistently.  To this
end, we will consider a theory arising from BiMOND~\cite{Milgrom:2009gv}.
 Additionally readers should consider the cosmologies of these theories,
 as suggested in~\cite{TimTom}.  

% A more thorough investigation of the weak field limit in these theories is left for future work~\cite{alibimetric}.

The structure of this paper is as follows, we firstly consider analytical
and numerical solutions from solving this theory around the Earth-Sun SP.  Next we move onto consider the integrated signal to noise ratio (SNR) as a measure of how sensitive the LPF instruments will be, the main results of which are contained in Figures \ref{fig:SNR cont} and \ref{fig:SNR null typeII}.  A look at converting a null signal into a constraint on the parameter space of our theories is  the subject of Section\ref{sec:SNR} - seeing whether we could differentiate between different modified gravity theories.  Finally Section\ref{sec:PFF} looks at parameterised approaches to these theories and looks to the future.  We leave detailed derivations and computational methods to the appendices.

%==============================================================================

% So far we have concentrated on type I non-relativistic MONDian theories,
% developing analytical and numerical predictions for the LPF saddle test.  It is interesting, however, in light of the results presented in Section \ref{theory} to consider whether the phenomenology associated with these theories are carried over into type II theories and whether there are any easy ways of determining between different theories potentially from data.

\subsubsection{Solutions, Tidal Stresses and Signal to Noise Ratios}\label{sec:SNR}

\begin{figure}[t!]\begin{center}
\resizebox{\columnwidth}{!}{\includegraphics{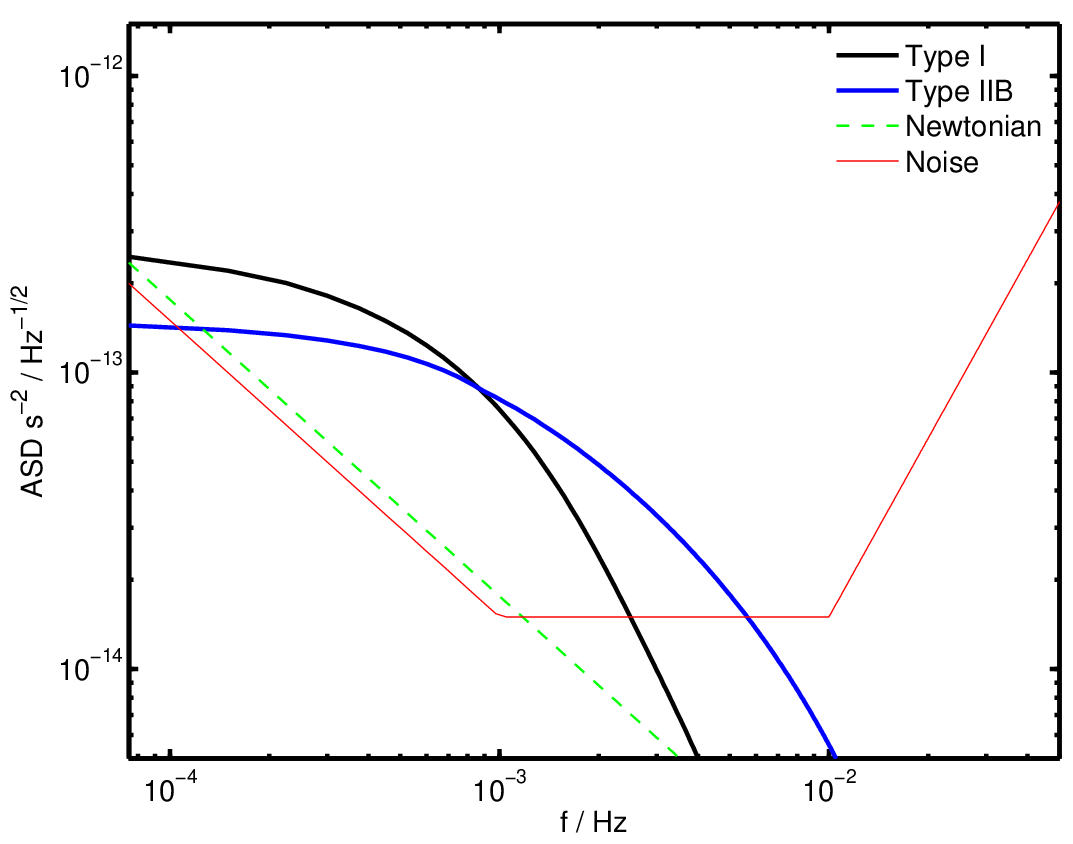}}
\caption{\label{fig:signal noise ASD}{ASD plot of the MONDian and rescaled
Newtonian signals, along with the noise profile.  Using the parameters $b = 50$km fly-by and $v = 1.5$ ms$^{-1}$, we compare between type I and
IIB theories, finding in this scenario SNRs of 28 and 35, respectively.}}\end{center}
\end{figure}

\begin{figure} \begin{center}
 \resizebox{\columnwidth}{!}{\includegraphics{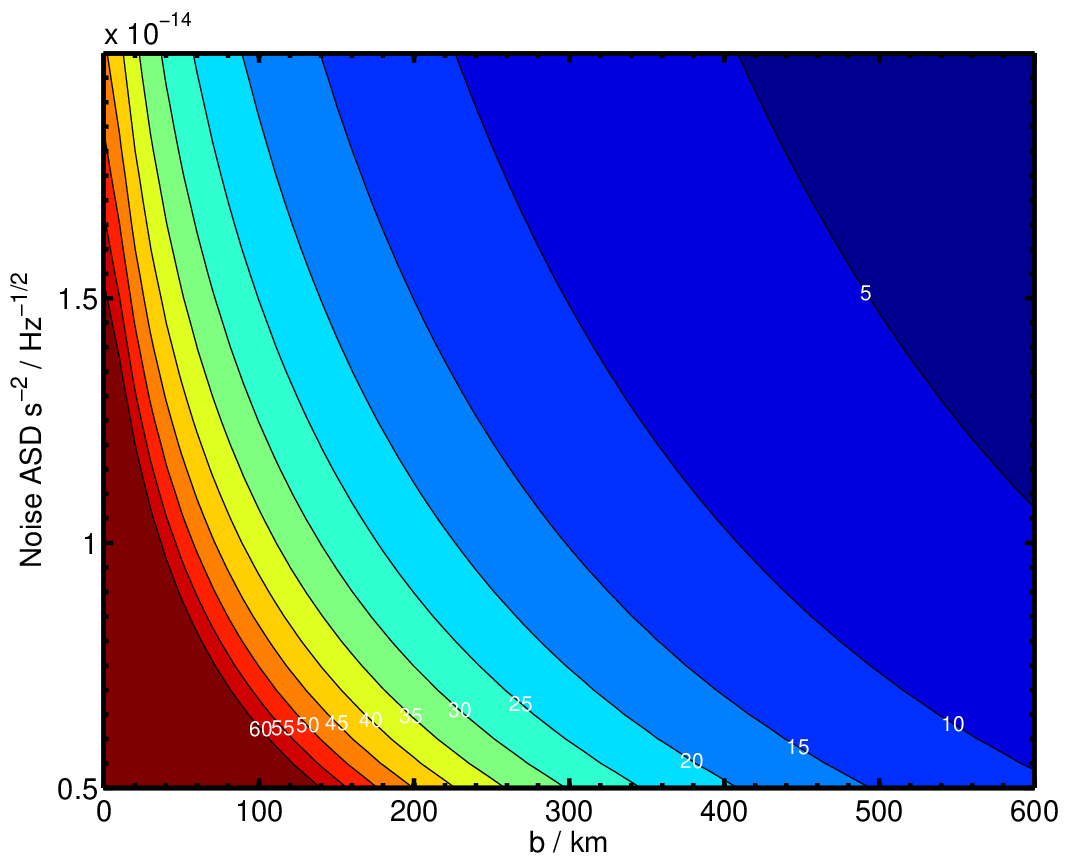}}
 \resizebox{\columnwidth}{!}{\includegraphics{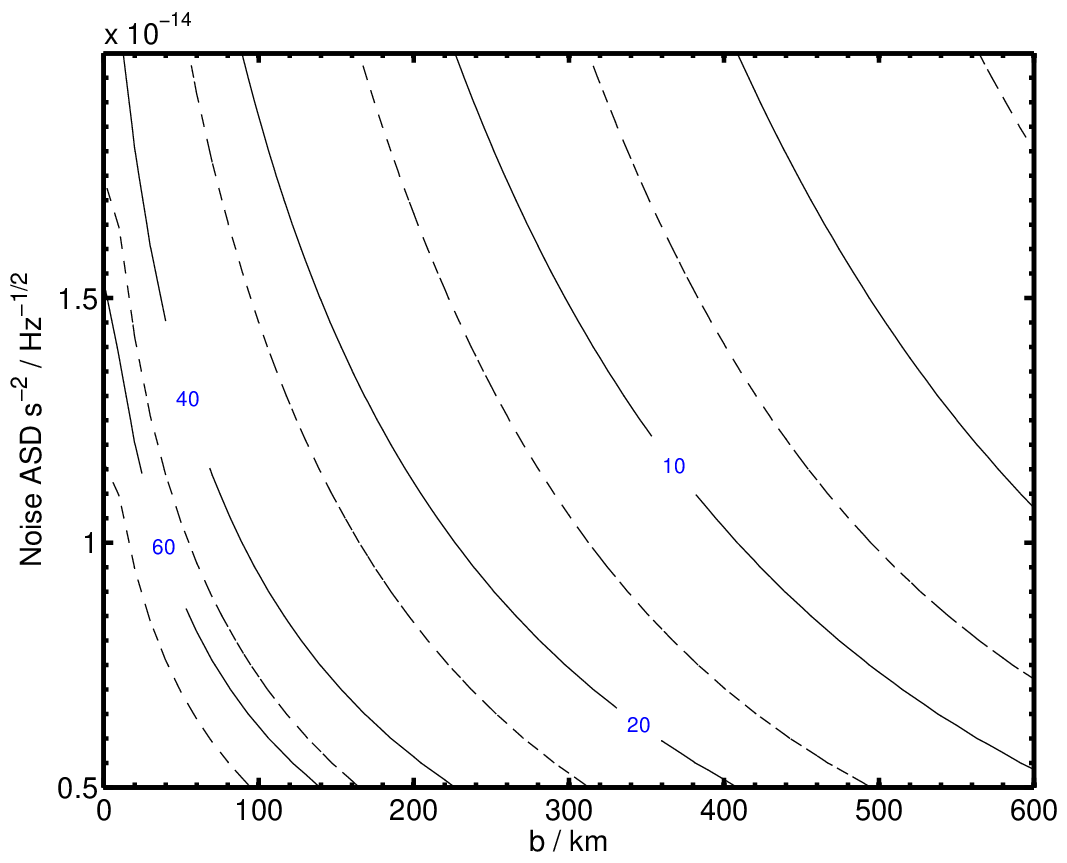}}\end{center}\caption{\label{fig:SNR cont}{{\bf Top panel}, SNR contours for various impact parameters and baseline ASD noise, for $v = 1.5$ kms$^{-1}$.  Calamitous assumptions would still lead to SNR in excess of 5. More optimistic ones (b around 50km or less, noise half way up the scale) would lead to SNRs easily around 55.  {\bf Bottom panel}, a comparison of SNR contour lines between type I and IIB theories.  The solid lines are the typical SNR to be obtained in IIB theories and the dashed lines to their immediate left the corresponding type I line - as we see IIB beats I.}}\end{figure}

In this analysis, we will make use of both analytical and numerical results.
We suggest using a free function $\nu$ of the form \be \nu = \left( 1 + \frac{1}{w^2}\right)^{1/4}\ee  We outline in Appendix \ref{typeIIapp} some reasons for this choice and
 with it we solve Equation ?? around the SP (recalling 
  we are free of sources),  \be \Del^2 \phi = \frac{\kappa}{4\pi} \Del\cdot(\nu\Del\Phi_N)
= \frac{\kappa}{4\pi} \left(\nu \underbrace{\Del^2 \Phi_N}_{= 0|_{SP}} +
\,  \Del \nu \cdot \Del \Phi_N\right)\nonumber \ee  We can approximate the
Newtonian near to the SP with a linear profile, which in spherical coordinates takes the form, \bea \vF_N = -\Del\Phi_N
&=& A r \vN = A r (N_r \ve_r + N_\psi \ve_\psi)\label{linearNewt} \nonumber\\ \nonumber N_r &=& \frac{1}{4}(1 + 3 \cos 2\psi)\\ N_\psi &=& -\frac{3}{4} \sin 2\psi\eea  Using this form of the Newtonian field as our source, we
proceed to solving the resulting system of equations.  Recall that the we
have both a deep MONDian (DM) regime, close to the saddle where $w \ll 1$ and an quasi-Newtonian (QN) regime further out, where $w \gg 1$.  Separating
these will be a boundary which using the linearised Newtonian we can find,
\bea w = \left(\frac{\kappa}{4\pi}\right)^2\frac{A r |\vect{N}|}{a_0} &=& 1\\ r^2 \left(\cos^2 \psi + \frac{1}{4}\sin^2\psi\right) &=& r_0^2 = \left(\frac{16\pi^2 a_0}{\kappa^2 A }\right)^2\eea As such we find an ellipsoidal (or bubble
shaped) boundary, with semi-major axis which we denote $r_0$.  Solving
the system, for the particular choice of $\nu(w)$, we have for the inner bubble, \bea -\Del\phi &=& \frac{4 \pi a_0}{\kappa} \left(\frac{r}{r_0}\right)^{0.5}
(F_r \vect{e}_r + F_\psi \vect{e}_\psi) \nonumber\\ F_r &\approx& 0.0354 + 0.2829\cos 2\psi - 0.0162\cos 4\psi \nonumber\\ F_\psi &\approx& - 0.3772\sin 2\psi - 0.0432\sin 4\psi  \eea and similarly for the outer bubble \bea -\Del\phi &=& \frac{4 \pi a_0}{\kappa} \left[\frac{r_0}{r}
(F_r \vect{e}_r + F_\psi \vect{e}_\psi) + \frac{r}{r_0}\vN \right]\nonumber\\ F_r &=& 1 \nonumber\\ F_\psi &\approx& - 0.5752 \sin 2\psi + 0.4652 \sin
4 \psi \eea
(The full calculational details are left to Appendix \ref{analytical}).  
\begin{figure} \begin{center}
\resizebox{1.\columnwidth}{!}{\includegraphics{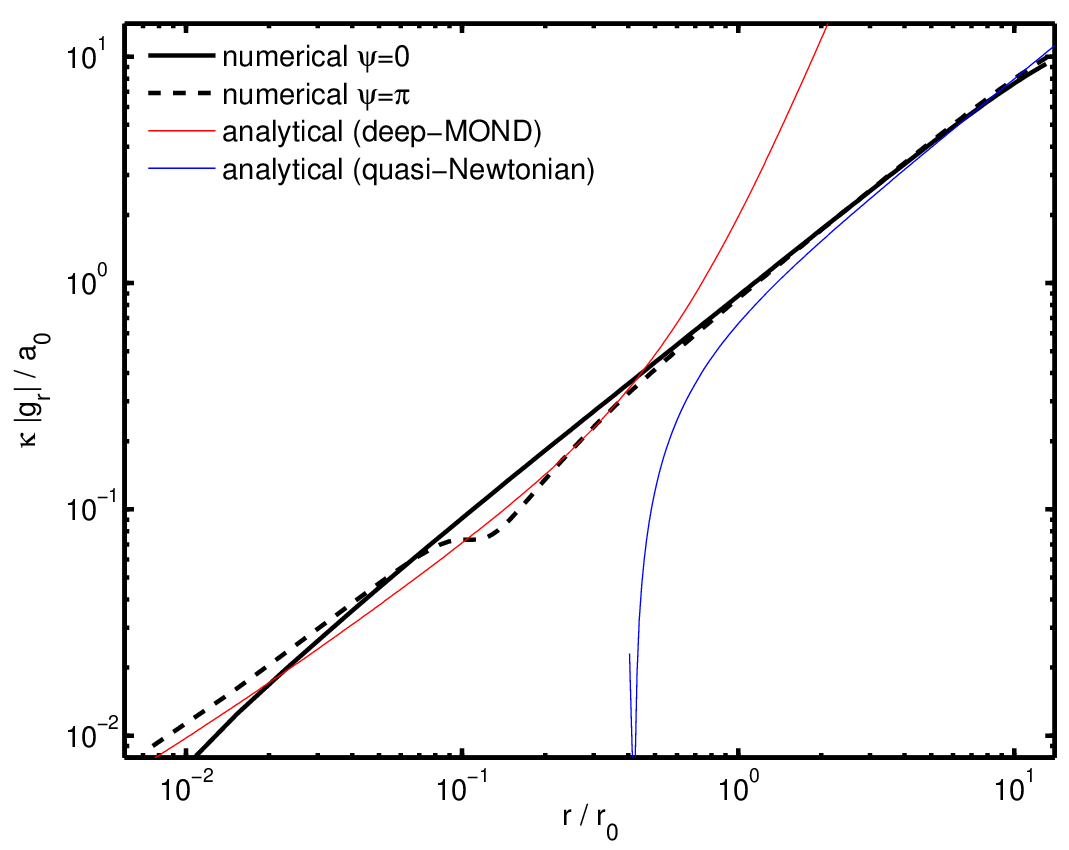}}
\caption{\label{fig:anal num solns}{A comparison between the numerical and analytical results for a component of $\vect{g} = -\Del\phi$ around the Earth-Sun saddle.}}\end{center}
\end{figure}
Now we can consider numerical solutions, using a relaxation code to solve (\ref{type II Laplacian}) on an adaptive lattice.  Such a code was first employed to study these theories in~\cite{bevis} and subsequently adapted
for more general free functions in~\cite{aliscaling}.
 In Appendix \ref{code}, we detail the modifications required for this work.  Consider the Earth-Sun saddle, using a $257^3$ lattice of physical size $10^4$ km with a typical central resolution of $\sim$2.6 km.  Running the code delivers us predictions for the modified forces close to the saddle, which as Figure \ref{fig:anal num solns} shows are a good match to analytical solutions in their respective domain and provide the appropriate interpolation in the intermediate region.  Next consider that LPF is sensitive to differential acceleration, from which we can infer the tidal stresses of our signal.  Note also that the $\phi$ field produces both a MONDian effective field and a rescaled Newtonian component.  Taking these into account, the observable tidal stress will be of the form \be S_{ij} = -\frac{\partial^2 \phi}{\partial x_i \partial x_j} + \frac{k}{4\pi}\frac{\partial^2 \Phi^N}{\partial x_i \partial x_j}\ee It is important that we know both field components to the same degree of accuracy or systematic errors can crop up due to the imperfect subtraction of the Newtonian field.  Using the techniques of noise matched
filtering from gravitational wave searches~\cite{sathya}, we can compute
the Signal to Noise Ratios (SNR) of these anomalous tidal stresses with LPF.  The basic idea laid is to correlate a time series $x(t)$ with an optimized template designed to provide maximal SNR, given the signal shape $h(t)$ and the noise properties of the instrument.  Whilst the specifics of the instrument noise won't be known properly until the satellite is {\it in situ}, there exist nominal requirements that it must meet~\cite{LPF1}, as well as best estimates for the noise signal waveform.  In our setup, we align the line joining the Sun-Earth
as our $x$ axis, such that we have trajectories of the form \be h(t)=S_{yy}(vt,b,0)\ee where $v$ is the velocity of the spacecraft, $b$ is the impact parameter and $t=0$ corresponds to the point of closest saddle approach. In a more general setup, for an approximately constant velocity $\bf v$, a closest approach vector $\bf b$, and with the masses aligned along unit vector $\bf n$, \be \label{hoft} h(t)=n^i n^j S_{ij}({\bf b}+{\bf v}t )\ee  The maximal SNR, realised by correlating the optimal template with the noise, is found to be \be \label{SNR}\rho_{\mathrm{opt}} = 2 \left[ \int_{0}^\infty \, df\frac{ \left |\tilde h(f)\right|^2 }{S_h(f)} \right]^{1/2} \ee where $\tilde{h}(f)$ is the Fourier transformed signal waveform and $S_h(f)$ is the noise waveform in Fourier space.  To make this analysis concrete, consider a flyby at $b = 50$km.  We plot the noise and signal, terms of amplitude spectral density (ASD), which is simply square root of power spectrum, in Figure \ref{fig:signal noise ASD}.   As we see, there is ample signal compared to idealised noise (here with baseline $1.5\times10^{-13}$ s$^{-2}/\sqrt{\textrm{Hz}}$).  This scenario gives an SNR of $35$ and for comparison a type I signal is presented
giving an SNR of 28 - note the differences in the behaviour at low and high frequencies.  The larger SNR here stems from the Poisson equation being linear in $\Del\phi$ and hence there are no associated curl forces (which generally exist in non-linear theories).  These will produce a softening of the tidal stress divergence, since this is not present here, we find a larger signal.  We produce in Figure \ref{fig:SNR cont} contours for typical SNRs produced at various impact parameters and baseline noises, as well as present a comparison of contours
from different theories - observing that type IIB beats type I on SNR.  As
Figure \ref{fig:signal noise ASD} shows, the position of the signal in fourier space is exactly where noise is lowest.  This highlights an uncanny coincidence, given that the accelerometer aboard LPF has a non-white noise profile, dipping in the region of the mHz (on the rough time scale of minutes). The motivation for such a design lies in the gravitational wave signals to be targeted by LISA - it just happens that the MONDian bubbles of anomalous tidal stresses around the Earth-Sun-Moon saddles are of length scale $\sim 10^3$ km and free-falling bodies around this region have a typical speed of $\sim 1$ km s$^{-1}$.  Put together, this suggests the time scale for crossing a MONDian bubble would be on the order of minutes - right where the instrument performance is optimal. 

\subsubsection{Designer $\nu$ functions}
It is very easy to construct free functions which mimic parameterised galactic
$\tilde{\mu}$ functions~\cite{Zhao}, for instance \be \nu = \frac{w^{-1/2}}{1 + \frac{4\pi \alpha}{\kappa} w^{n-1/2}} \ee giving the usual $\nu \rightarrow 1/\sqrt{w}$ in the DM regime but moving to a different power law, $\nu \sim 1/w^{n}$ (tunable by the value of $\alpha$) for larger accelerations.  We can attempt the exercise of designing a free function, based on a null result (taking an upper bound from a SNR = 1 result) at some acceleration.  We can then convert this into a restriction on the $\nu$ parameter space - clearly
the SP bubble clearly will have to shrink.  We start by fixing the asymptotica, the astrophysical regime gives us $\nu \approx 1/\sqrt{w}$ for $F_N \leq a_0$.  Far from the SP, we will have the $\nu \approx 1$ regime, but in the intermediate regime between the two (which we will be probing),
lets suggest a model such as: \bea \nu\approx 1/\sqrt{w}&\quad {\rm for}\quad&
 w<\left(\frac{k}{4\pi}\right)^2\\
\nu\approx \left(\frac{w^{trig}}{w}\right)^n
&\quad {\rm for}\quad &\left(\frac{k}{4\pi}\right)^2<w<w^{trig}\\
\nu\approx 1&\quad {\rm for} \quad &w>w^{trig}
\eea
where the point when non-Newtonian behaviour in $\phi$ is triggered can be interchangeably pinpointed by:
\bea
w^{trig}&=&\left(\frac{\kappa}{4\pi}\right)^{2-\frac{1}{n}}\\
a_\phi^{trig}&=&a_0\left(\frac{\kappa}{4\pi}\right)^{1-\frac{1}{n}}\\
a_N^{trig}&=&a_0\left(\frac{\kappa}{4\pi}\right)^{-\frac{1}{n}}
\eea
We still have that when $a_N<a_0$, the field $\phi$ dominates $\Phi_N$ - as per our requirements.  However now the intermediate region where $\phi$ hasn't yet dominated but is already non-Newtonian is in a narrower band of accelerations $a_0<a_N<a_N^{trig}$. As a result, the MOND bubble shrinks in this model according to  \be r_0\approx 383 \left(\frac{\kappa}{4\pi}\right)^{\frac{2n-1}{n}} \unit{km} \label{tIIbubblesize}\ee This result shows that for a given null measurement up to some acceleration $a_N^{trig}$, using this general argument, our constraints between type I and IIB theories will be different.  In this case, the bubble size would be expected to shrink more than in the type I case (where the exponent is just $\frac{n-1}{n}$~\cite{ali}), given the sharper divergence in the tidal stress (and so larger signal) and this is exactly what (\ref{tIIbubblesize}) suggests.  Thus our naive expectation that the
bubble would be smaller, given the stronger signed expected, compared to type I theories turns out to be true.  \begin{figure} \begin{center}
\resizebox{\columnwidth}{!}{\includegraphics{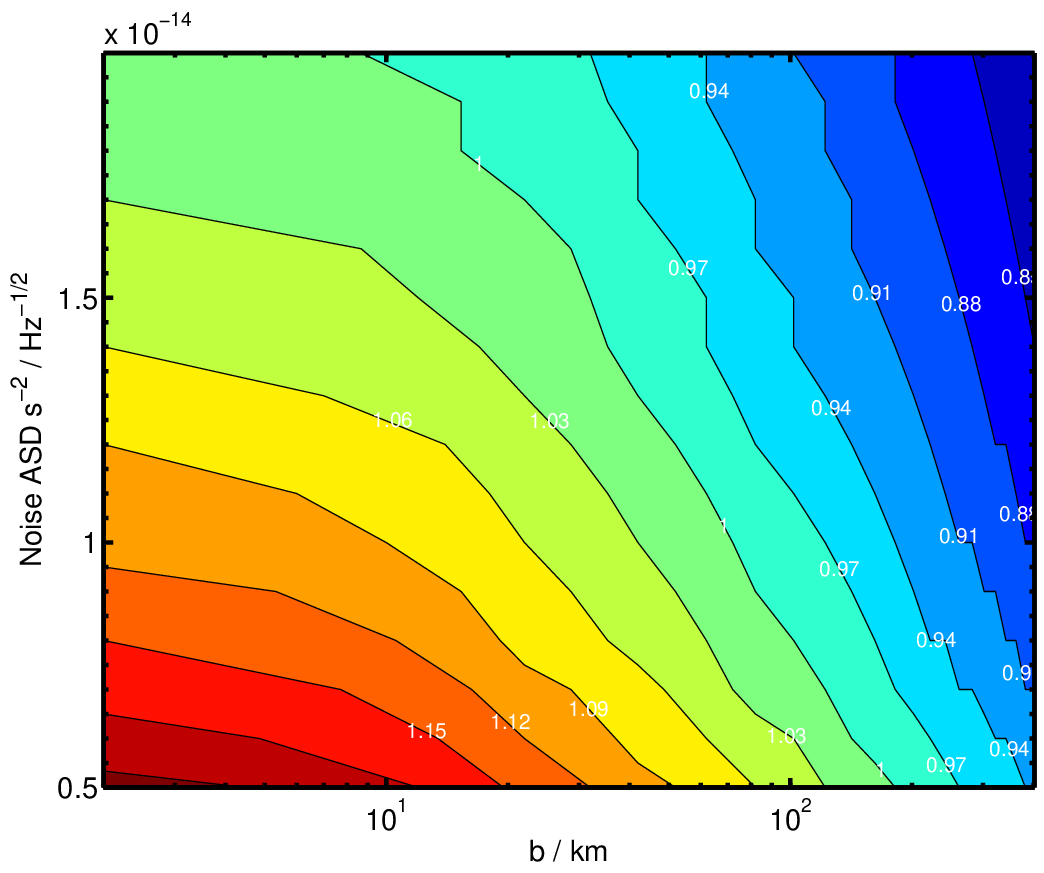}}
\caption{\label{fig:SNR null typeII}{Contours of the power $n$ needed to obtain SNR $= 1$ for different impact parameters and noise levels, upto $b = 400$ km using our designer free function.}}\end{center} \end{figure} 
One issue that we come up against here is that because of there are many
varying transients from $\nu \rightarrow 1$, making a model dependent statement is beyond our reach - quite simply because a null result only lets us probe the regime of $b\gg r_0(n)$.  Performing an order of magnitude argument
however is possible using a designer function, \be \nu(w) = 1 + \left(\frac{w^{trig}}{w}\right)^n\ee Assuming spherical symmetry \be \vF_\phi = \frac{k}{4 \pi}\, \nu \, \vF_N \label{nu grad phi sym}\ee breaking this up into a background contribution
and a modified part \be \vF_\phi={}^0\vF_\phi+\delta \vF_\phi \ee where ${}^0\vF_\phi=\frac{\kappa}{4\pi}\vF_N$. Substituting in and solving gives \be \delta \vF_\phi = \frac{k}{4 \pi}\, (\nu - 1)\,\vF_N \simeq \left(\frac{a_0}{|\vF_N|}\right)^n\vF_N \label{order of mag null} \ee  from which the tidal stresses can be inferred.  Figure
\ref{fig:SNR null typeII} shows the resulting values of $n$ required for a SNR = 1 result.  As we see, the dynamics here are somewhat different which opens up the future possibility of letting us differentiate between type I and IIB theories.  A result in one theory could be considered ``unnatural'' but could possibly viable in another.  

%================================================================================
\subsubsection{Parameterised Free Functions}\label{sec:PFF}
\begin{figure} \begin{center}
\resizebox{\columnwidth}{!}{\includegraphics{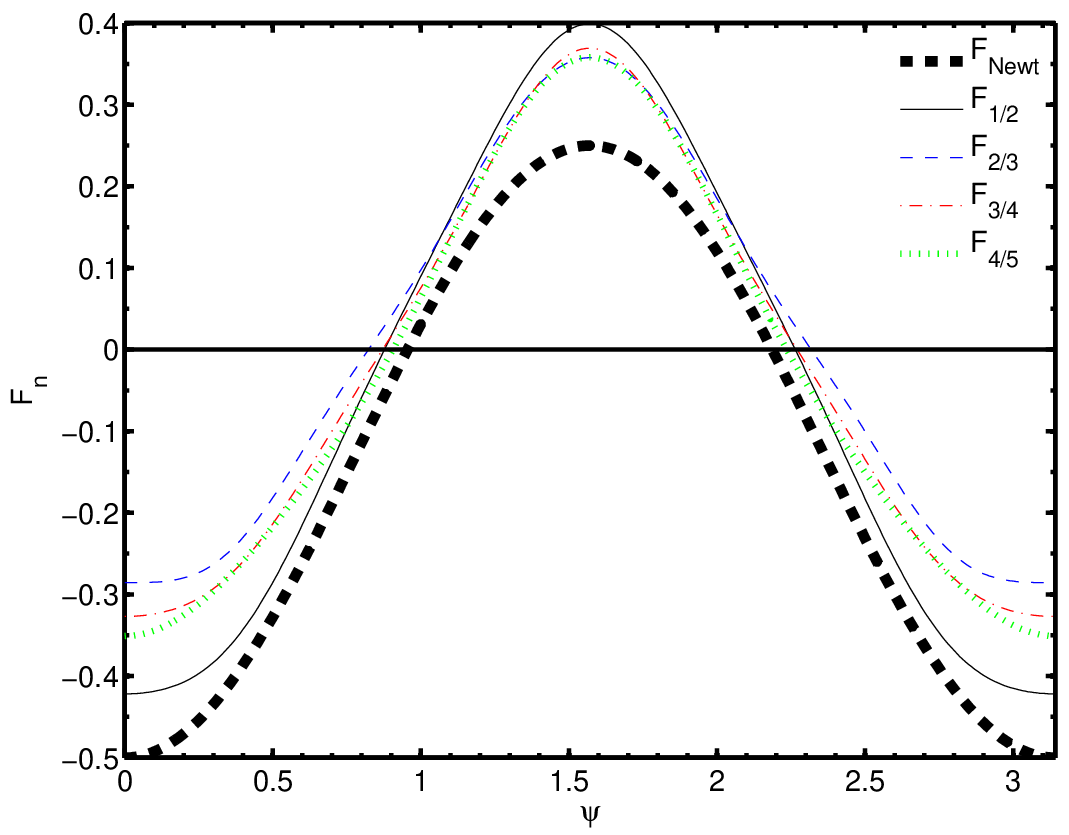}}
\resizebox{\columnwidth}{!}{\includegraphics{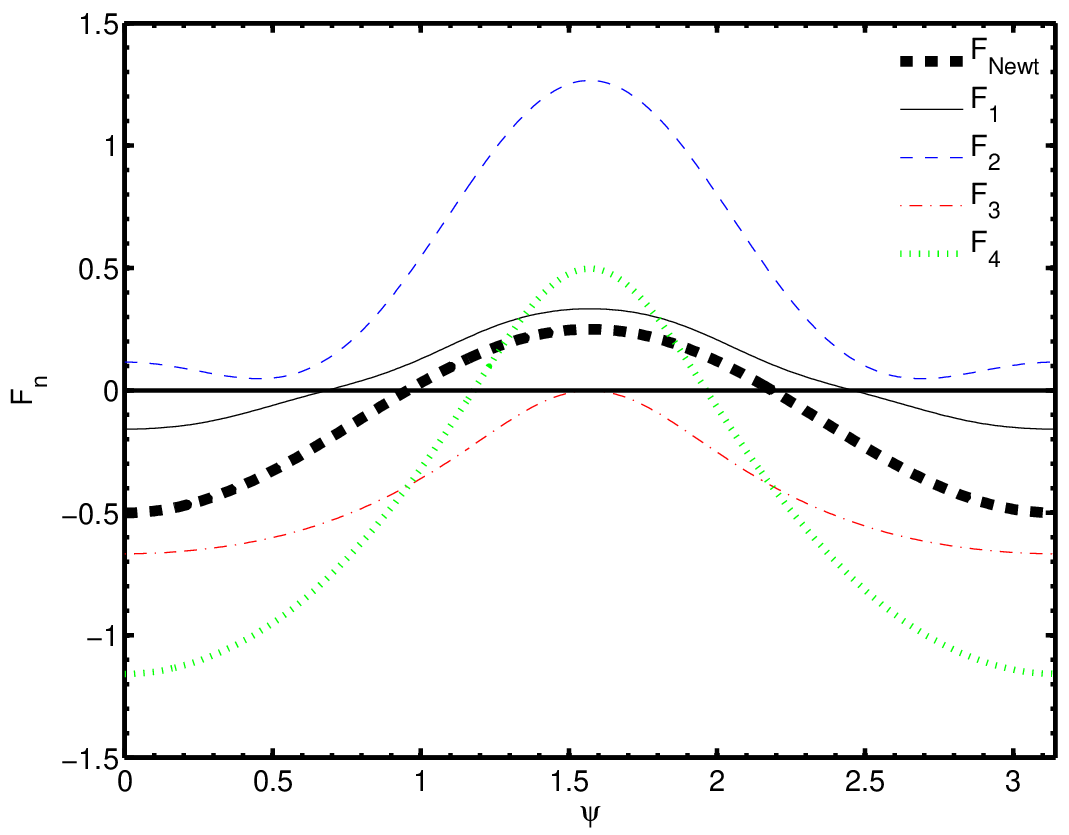}}
\caption{\label{fig:typeIIgeneralproifles}{The angular profile functions F for the inner ({\bf top panel}) and outer ({\bf bottom panel}) bubble regions, alongside the linear Newtonian potential profile $F_{Newt}$.  The values
of $a,b$ are used to label each function.  The results clearly show the relative invariance of the inner bubble potential profile functions compared to the significant differences for the outer bubble.  }}\end{center} \end{figure}

Other questions to be asked include what inferences can we make from data if LPF detects anything (after ruling out systematics and noise).  Is it possible to pick out particular features of our free functions?  Would there be a way to potential discriminate between different types of theory?  Here we seek to address some of these briefly.  Firstly, lets develop a parameterised $\nu(w)$, which will act as a prototype for further discussions, \be \nu = \left(\frac{1+ w^b}{w^b}\right)^{a/b}\ee notice that in the requisite limits \bea w \ll 1, \,\,\,\,\,\, \nu &\simeq&
\frac{1}{w^a} + \dots \\ w \gg 1, \,\,\,\,\,\, \nu &\simeq& 1 + \frac{a}{b}\frac{1}{w^b} + \dots \eea This reduces back to MOND for the case of $a = {1\over2}$ and $b \geq 1$, but otherwise opens up the parameter space.  Equation {type II Laplacian} has the form \bea \Del^2 \phi &=& \frac{4 \pi a_0}{\kappa}\frac{a\,r_0^{b-1}}{r^b N^{b}}\left(N_r + \frac{\partial N}{\partial \psi} \frac{N_\psi}{N}\right)\left(1 + \left(\frac{r_0}{r N}\right)^b \right)^{a/b-1} \nonumber\\ &\simeq& C_1 r^{c-2} S_c(\psi) +
\dots \eea An ansatz for the leading order contribution takes the form \bea \phi &=& C_1 r^c F(\psi)\eea with the profile function $F(\psi)$ satisfying the ODE \be c(c+1) F + F' \cot \psi + F'' = S_c(\psi) \label{generalODE}\ee Note this satisfies the equations $b \neq 2$ and we will consider
the $b = 2$ case separately later.  The constant and source terms are given by \bea  C_1 &=& a\frac{4 \pi a_0}{\kappa} r_0^{c-3}  \\ S_c(\psi) &=& \frac{1}{N^{c-2}}\left(N_r + \frac{\partial N}{\partial \psi}\frac{N_\psi}{N}\right) \label{generalsource}\eea such that in each regime \bea
c = 2 - a  && w \ll 1\\ c = 2 - b && w \gg 1\eea With these in mind, we can
solve the inhomogenous ODE, as before, to find the variation in profile functions
for different $a,b$.  To make a connection with other modified force laws~\cite{aliscaling},
consider under spherical symmetry: \bea \left(\frac{\kappa}{4\pi}\frac{F_\phi}{a_0}\right)^n F_\phi = \frac{\kappa}{4\pi}F_N \Rightarrow F_\phi &=& \left(\frac{\kappa}{4\pi}\right)^{\frac{1-n}{1+n}}
F_N^{\frac{1}{n+1}} a_0^{\frac{n}{n+1}} \nonumber\\ F_\phi = \left(\left(\frac{\kappa}{4\pi}\right)^2\frac{F_N}{a_0}\right)^{-a}\frac{\kappa}{4\pi}F_N
&=& \left(\frac{\kappa}{4\pi}\right)^{1-2a}F_N^{1-a}a_0^{a}\nonumber\\\eea and so let us, without loss of generality, parameterise \be a \rightarrow \frac{n}{n+1} \ee  This means the choice of $n=1 \Leftrightarrow a
= 1/2$ gives the required force law $F_\phi = \sqrt{F_N a_0}$ and some $n
> 1$ will parameterise deviations from it.  We illustrate different choices of $b,n$ in Figure \ref{fig:typeIIgeneralproifles}.  The take home message from this analysis is that in the strongly modified regime of the inner SP bubble, there is little variance between the potentials - the main difference is from the radial exponent.  This means one can expect the magnitude of any resulting anomalous tidal stresses to be different rather than the shape of the time series.  This has an impact on the noise matched filtering techniques of Section\ref{sec:SNR}, as the space of inner bubble templates can effectively be significantly reduced.  With this generalised approach, consider the tidal stresses $S_{yy}$ computed by employing a change of variables \bea S_{yy} &=&  C_2\, r^{c-2} \,((c-2) \cos 2\psi - c)\, c F  \nonumber \\ &-& \left. 2\cos \psi \left(2 (c-1) \sin \psi \,F'+\cos \psi \,F''\right)\right)\eea where $C_2 = C_1 / 2$.  Although slightly complicated, this expression demonstrates that in the regime where the linear Newtonian \ref{linearNewt}  can be applied, there is a neat separation of variables between radial and azimuthal functions.  Notice that in all these models, the stresses will be divergent since $c < 2, \forall a,b > 0$.  For the $b = 2$ case,  \bea \phi &=&\frac{4\pi a_0}{\kappa} \left( r_0\ln \left(\frac{r}{r_0}\right)
+ H(\psi) \right)\\ S_{yy} &=& -\frac{4\pi a_0}{\kappa r^2} \left(r_0 \cos 2\psi - \sin 2\psi F'+ \cos^2 \psi F''\right) \nonumber \\\eea where $H(\psi)$
is a function solved by the second order ODE (\ref{generalODE}) sourced by
(\ref{generalsource}).  The main reasoning behind the invariance of profile functions in the low acceleration regime and the Sm\"org\aa sbord of solutions in the other stems from the behaviour of the source term \be \lim_{\psi \rightarrow \pi/2} S_c = -2^{c-3}\ee For $0 < a < 1$, $S_c$ remains relatively unchanged compared to $b > 0$ where $S_c$ becomes increasingly singular at $\psi = \pi/2$ for increasing $b$.  Observe that $S_{yy} \sim r^{-\#}$ where $\nu \rightarrow w^\#,w \ll 1 $ or $\nu \rightarrow 1 - \frac{C}{w^\#} + \dots,
w \gg 1$ whereas in the type I case~\cite{aliscaling}, $S_{yy}$ was found to scale much more conservatively in the DM regime.  

%================================================================================
\subsubsection{Conclusions}\label{concs}
Our investigations have shown how a LPF saddle flyby could either detect
modified gravitational behaviour to a high SNR or place stringent constraints
on it.  With an appropriate noise model, SNRs $\geq 35$ could
be expected missing the saddle by 50km or less (a larger figure than would
be expected from type I theories).  The question then arises as to how generic this conclusion is, or conversely, should a negative result be found what
can be said about these theories.  Predictions for what happens inside the bubble are found to be model independent, the tidal stress anomalies outside the bubble however do depend on the transient from the modified into the
Newtonian regime, with a model dependent fall-off.  Thus for impact parameters smaller than $r_0$ the predicted SNRs are robust and do not change substantially
- the currently expected case is $b \leq 10\unit{km}$ (with $r_0\sim 383\unit{km}$).  A way therefore for such theories to wriggle out of a negative LPF result would be to change the bubble size.  This could be accomplished
with a ``designer'' function - giving the free function two scales and two power-laws before reaching the rescaled Newtonian regime.  Furthermore the effects inside the bubble are different from type I predictions, but generically stronger. This stems from type II theories lack of curl field, a feature which appears to soften out the anomalous stresses.  This results generically in larger SNRs between type I and IIB theories.  As we see applying the same arguments between the two theories results in different constraints and the bubble size generically being smaller here compared to type I for a null result.  By considering a generalised approach to free functions,
we find inner and outer bubble solutions and considered the scaling of the tidal stress with parameters extracted from the free functions.  The main result being that in the DM regime, our angular profile functions become relatively invariant between models, making the prospect of extracting features of $\nu$ from potential data interesting - the main contribution being the radial parameter and its associated exponent.  The divergence of the tidal stresses scale much quicker for a given $\nu(a,b)$ compared to the corresponding $\mu(a,b)$ models of type I theories.  These models suggest that tying a
result to either type I or IIB theories is in principle viable - differentiating
between the two theories is a realistic prospect, our imagination is only
limited to the theories we develop.

%=============================================================================
\begin{acknowledgments} 
The author would like to thank Jo{\~a}o Magueijo, Tim Clifton, Johannes Noller and Dan Thomas for useful discussions and suggestions.  The author thanks the STFC, Department of Physics and Centre for Co-Curricular Studies, Imperial College for financial support during the various stages of this
work.  Numerical work was carried out on the COSMOS supercomputer, which is supported by STFC, HEFCE and SGI.
\end{acknowledgments}

%\newpage
\appendix

%=============================================================================
\section{Classifying MONDian theories}\label{theory}

Our job is to approach theories where such modified behaviour is present and see if they represent good prospects for detection.  Their complexity and differences arise from the requirement that they should explain relativistic phenomena (such as lensing and structure formation) without appealing to dark matter, whilst in the non-relativistic regime have some Newtonian and
other modified limit. The manner in which such effects are manifest may however vary widely and there have been many previous studies as to the phenomenology of these ideas, particularly in this non-relativistic regime~\cite{typeIIpaper,ali,Milgrom:2009gv,zlosnik:2006aug,teves}.
 We will briefly outline some of these here, with the caveat that this list
 is neither exhaustive, nor represents the final story on gravity theories
 at the time of writing and for a more in depth look at gravity theories,
 we point the reader towards~\cite{clifton:2012mar}.
 
 \begin{itemize}
\item{\bf Type I:} Here the total potential acting on non-relativistic particles
is given by the sum of the usual Newtonian potential $\Phi_N$ and a fifth force field, $\phi$: \be \Phi = \Xi\Phi_N+\phi\ee where $\Xi$ is some constant
usually set to unity and the Newtonian potential satisfies the usual Poisson equation \mbox{$\nabla^2 \Phi_N=4\pi G \rho$}, and the field $\phi$ is governed by: \be \nabla \cdot \left(\mu(z)\nabla \phi\right) = \kappa G \rho \label{type1} \ee The argument of $\mu(z)$ is given by \be z=\frac{\kappa}{4\pi}\frac{\vert\nabla\phi\vert}{a_0} \ee where $\kappa$ is a dimensionless coupling constant.  $\mu$ is a free function,
typically chosen limits of the theory are $\mu\rightarrow 1$ when $z\gg 1$ and $\mu \simeq z$ for $z\ll 1$.  The effect of these fields is twofold, in the large $z$ regime, $\phi \rightarrow \frac{\kappa}{4\pi} \Phi_N$ mimicking the Newtonian, this makes the physical potential have the form \be \Del \Phi \rightarrow \left(\Xi + \frac{\kappa}{4\pi}\right)\Del\Phi_N\ee
or equivalently the form of Newton's constant is altered \be G_{ren} \rightarrow
\left(\Xi + \frac{\kappa}{4\pi} \right)G_N \ee Cosmology sets bounds on the
variation of $G$, from BBN and effects in the CMB~\cite{CarrollLim, Nconstraint}.
 
 Additionally these two fields mean that the Newtonian behaviour is always
 present in the non-relativistic regime and non-linear behaviour in $\phi$ gets triggered at a certain acceleration \be a_{\text{trig}} = \left(\frac{4\pi}{\kappa}\right)^2
a_0\ee The field however remains sub-dominant until $a_N = a_0$ and this
is when fully modified behaviour is seen (in the galactic regime).  It is
this onset of non-linearity that we hope to probe with LPF.

\item{\bf Type II:} These are similar in set-up to type {\bf I}, with $\Phi = \Phi_N +\phi$ and
$\phi$ governed by a driven linear Poisson equation:\be \nabla^2 \phi = \frac{\kappa}{4\pi}\nabla\cdot\left(\nu(w)\nabla \Phi_N \right) \label{type2} \ee The argument of $\nu$ is given by \be w = \left(\frac{\kappa}{4\pi}\right)^2\frac{\vert\nabla\Phi_N \vert}{a_0}  \ee Once again $\nu$ is a free function and typically we give it the form
$\nu\simeq 1/\sqrt{w}$ for $w\ll 1$ and $\nu \rightarrow \unit{constant}$
for $w\gg 1$.  \\ We divide this up in the subtypes of {\bf IIA} or{ \bf IIB} with qualitatively very different implications, which can we seen more clearly if we return to using the physical
potential form \bea \Del^2 \Phi &=& \Del \cdot\left( \hat{\nu} \Del\Phi_N\right)
\\  \hat{\nu} &=& 1 + \left(\frac{\kappa}{4\pi}\right)\nu\eea 
Consider in the large $w$ regime: \begin{itemize}
\item{In type {\bf IIA}, $\nu \rightarrow 0$ which implies no $G$ renormalisation occurs and $a_{\text{trig}} = a_0$.  The whole theory in fact hinges on $\Phi$, all other fields are considered auxiliary.}
\item{In type {\bf IIB}, $\nu \rightarrow 1$ means a trigger acceleration similar to type {\bf I}.  }\end{itemize}

\item{\bf Type III:} Crucially, here non-relativistic particles are sensitive to a single field $\Phi$, satisfying a non-linear Poisson equation: \be \nabla\cdot\left(\tilde{\mu}(x)
\nabla \Phi \right) = 4 \pi G \rho \label{type3} \ee where the argument of
$\tilde\mu$ is \be x = \frac{\vert\nabla\Phi\vert}{a_0}\ee so that $\tilde \mu\rightarrow 1$ when $x\gg 1$ and $\tilde \mu\sim x$ for
$x\ll 1$. Again no renormalisation of $G$ and a trigger acceleration
$a_{\text{trig}} = a_0$
\end{itemize}
As the trigger acceleration sets the scale of the SP bubble, using the current estimates for our parameters ($\kappa = 0.03, a_0
= 10^{-10}\unit{ms}^{-1}$) we find these to be \bea a_{\text{trig}} &=& \left(\frac{4 \pi}{\kappa}\right)^2 a_0 \simeq 10^{-5}\unit{ms}^{-2} \Rightarrow r_0\sim 383 \unit{km}\nonumber\\ a_{\text{trig}} &=& a_0 = 10^{-10} \unit{ms}^{-2} \Rightarrow r_0
\sim 2.2 \unit{m}\eea
These distinctions group together types {\bf I} and {\bf IIB} as the best candidates for detection with LPF; types {\bf IIA} and {\bf III} would easily escape any negative result.

% The most significant distinction between the non-relativistic limits listed here links type I and IIB theories in opposition to type IIA and type III theories.  In the former, non-relativistic particles are sensitive to two fields, which mimic each other in the Newtonian regime and so the the gravitational constant is effectively renormalised. In the Newtonian regime (non-relativistic limit, with large total Newtonian force), we have $\mu\approx 1$ or $\nu\approx 1$, and so $\phi$ becomes proportional to $\Phi_N$: \be\label{rat} \phi\approx \frac{\kappa}{4\pi}\Phi_N\ \ee which has the effect of renormalising the observed gravitational constant \be G_{Ren}\approx G{\left(1+\frac{\kappa}{4\pi}\right)}\ee and $G_{Ren}$ is the gravitational constant measured, say, by the Cavendish experiment.  Nevertheless cosmology is sensitive to the bare $G$ (for example the Friedmann equations) and constraints arising from Big Bang nucleosynthesis and the cosmic microwave background (CMB)~\cite{CarrollLim, Nconstraint} fix $\kappa$ to be of the order of $10^{-2}$ or smaller and structure formation considerations may further fix it (see~\cite{kostasrev}). The conclusion being that in the non-relativistic regime, the field $\phi$ must be suppressed when $a_N=|\nabla\Phi_N|$ is much larger than $a_0$.

An important distinction here stems from the fact that
we have a curl term (often called a magnetic field) in type {\bf I} and {\bf
III} theories.  This is easiest seen when one attempts to linearize the non-linear Poisson equations present by introducing an auxiliary vector field (e.g. $\mu\nabla\phi$ for type {\bf I} theories) - such a field has non-zero curl. The same is not true for type {\bf II} theories, being already linear in $\phi$ and driven by a function of the Newtonian field, $\nu\nabla \Phi_N$, (a quantity which has a curl). This turns out to have a significant quantitative effect upon the magnitude of the saddle tidal stresses, as the magnetic field is known to soften the anomalous tidal stresses around the saddle points in type {\bf I} theories. 

\section{Analytical Arguments}\label{analytical}
%===================================================================

%\subsection{Type II Free Functions}
\label{typeIIapp}
%\subsection{Analytical Solutions}
\label{anal solns}
Our plan will be to consider functions similar to those investigated for
type I theories and so easily compare between the two.  We start with the
idea that {\it under the assumption of spherical symmetry}\bea \nonumber \Del \cdot (\mu\,\vF_\phi) = \frac{\kappa}{4\pi} \Del\cdot\vF_N &\Rightarrow&
\mu \vF_\phi = \frac{\kappa}{4\pi} \vF_N \\ \Del\cdot \vF_\phi = \frac{\kappa}{4\pi}\Del\cdot(\nu \vF_N) &\Rightarrow& \vF_\phi = \frac{\kappa}{4 \pi} \nu \vF_N \eea with
the natural comparison between \be \mu(z) \longleftrightarrow \frac{1}{\nu(w)} \nonumber\ee We will be inspired by the form of the type I free function $\mu$ considered previously~\cite{bekmag,bevis,aliscaling}, which satisfied \be\frac{\mu}{\sqrt{1 - \mu^4}} = z\ee where $z = \frac{\kappa}{4\pi}\frac{|\Del\phi|}{a_0}$.
 In the $z \gg 1$ limit, \be \mu \simeq 1 - \frac{1}{4z^2} + \dots\ee
which suggests in the analogous $w \gg 1$ limit, we need a function satisfying \be \nu \simeq 1 + \frac{1}{4 w^2} +\dots \label{nuexp}\ee
Using the definition of $z$, under spherical symmetry\bea z = \frac{\kappa}{4\pi}\frac{|\Del\phi|}{a_0} &=& \nu \left(\frac{\kappa}{4\pi}\right)^2\frac{|\Del \Phi_N|}{a_0}
= \nu w \\ \frac{\mu}{\sqrt{1 - \mu^4}} &=& \frac{1/\nu}{\sqrt{1+1/\nu^4}} = \nu w\eea which we can solve to find \be \nu = \left(1 + \frac{1}{w^2}\right)^{1/4} \ee and note in the quasi Newtonian regime, this mimics the behaviour of (\ref{nuexp}).  Whilst we stress this derivation is only strictly valid for spherical symmetry, it remains a good starting point for comparison between these theories.  Next we expand the
Poisson equation \be \nonumber\Del^2 \phi = \frac{\kappa}{4\pi} \Del\cdot(\nu\Del\Phi_N)
= \frac{\kappa}{4\pi} \left(\nu \underbrace{\Del^2 \Phi_N}_{= 0|_{SP}} +
\,  \Del \nu \cdot \Del \Phi_N\right) \ee and then use the linear Newtonian approximation, \bea \vF_N = -\Del\Phi_N
&=& A r \vN = A r (N_r \ve_r + N_\psi \ve_\psi)\label{linearNewtApp} \nonumber\\ \nonumber N_r &=& \frac{1}{4}(1 + 3 \cos 2\psi)\\ N_\psi &=& -\frac{3}{4} \sin 2\psi\eea Remembering that the characteristic MONDian bubble size, denoted $r_0$, is given by the expression \bea r_0 = \left(\frac{4\pi}{\kappa}\right)^2\frac{a_0}{A} \eea Such that we can write \bea  w = \frac{\kappa^2}{16 \pi^2}\frac{|\Del \Phi_N|}{a_0} = \frac{r}{r_0}N\eea
giving us the form of the source term  \bea \Del^2\phi = \frac{4\pi a_0/\kappa}{\sqrt{4
r r_0}}\left(\frac{N_r}{N^{1/2}} + \frac{N_\psi}{N^{3/2}} \frac{\partial N}{\partial \psi}\right)\left(1 + \left(\frac{r N}{r_0}\right)^2\right)^{-3/4} \label{sourcePoisson}\eea The problem here is akin to electrostatics, solving the equations subject to the boundary conditions that $\delta \vF_\psi$ vanishes (and $\delta \vF_r$ equate) at $\psi = 0$ and $\pi$, such that we avoid a jump in the field at $\psi = \pi/2$. 

\subsubsection{DM Regime}
 For $r \ll r_0$, it's clear Equation (\ref{sourcePoisson}) reduces to: 
%  \bea\Del^2\phi = \left(\frac{4\pi a_0}{\kappa}\right)&&\left(\frac{N_r}{N^{1/2}} + \frac{N_\psi}{N^{3/2}} \frac{\partial N}{\partial \psi} \right) \\\nonumber &&\left(\frac{1}{ 4 r\,r_0}\right)^{1/2}\left\{1 - \frac{3}{4} w^2 + \dots \right\}\eea

\bea\Del^2\phi = \frac{4\pi a_0/\kappa}{\sqrt{4 r N r_0}}\left(N_r + \frac{N_\psi}{N} \frac{\partial N}{\partial \psi} \right) \left\{1 - \frac{3}{4} w^2 + \dots \right\}\nonumber\\\eea where the angular functions of the
leading order term neatly reduce to \be \frac{7 + 9\cos 2\psi}{(2(5 + 3\cos 2 \psi))^{5/4}} = g(\psi)\label{RHS}\ee  The separable form of the source
suggests an ansatz of\be \phi = C_1 \,r^a \,F(\psi)\ee where $C_1$ is a constant,
the exact form of which is fixed from the source term.  This gives rise to a sourced second order ODE: \be r^{a-2} \left(a(a+1)F + \cot(\psi)\, F' + F'' \right) = r^{-1/2}\,g(\psi) \label{DM ODE}\ee where $' = \partial / \partial \psi$ and \be C_1 = \frac{4\pi a_0}{\kappa\sqrt{r_0}} \ee We find that the solutions of the homogenous equation are Legendre Polynomials of order $a$, with the form of (\ref{DM ODE}) suggesting $a = 3/2$ and the inhomogeous solution is found to be\bea  F \approx - 0.0236 - 0.1886 \, \cos 2\psi + 0.0108 \, \cos 4\psi \eea We can then compute the components of the
MONDian force \bea -\Del\phi &=& \frac{4 \pi a_0}{\kappa} \left(\frac{r}{r_0}\right)^{0.5}
(F_r \vect{e}_r + F_\psi \vect{e}_\psi) \nonumber\\ F_r &\approx& 0.0354 + 0.2829\cos 2\psi - 0.0162\cos 4\psi \nonumber\\ F_\psi &\approx& - 0.3772\sin 2\psi - 0.0432\sin 4\psi  \eea 
\begin{figure} \begin{center}
\resizebox{\columnwidth}{!}{\includegraphics{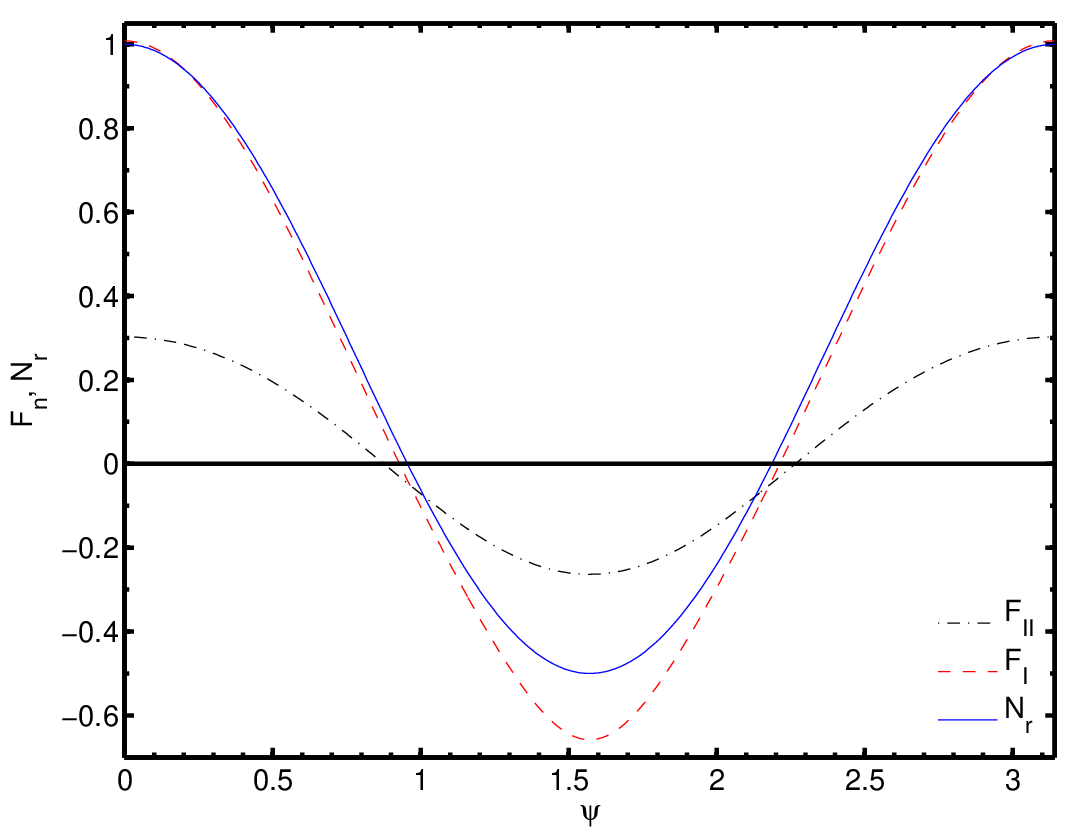}}
\resizebox{\columnwidth}{!}{\includegraphics{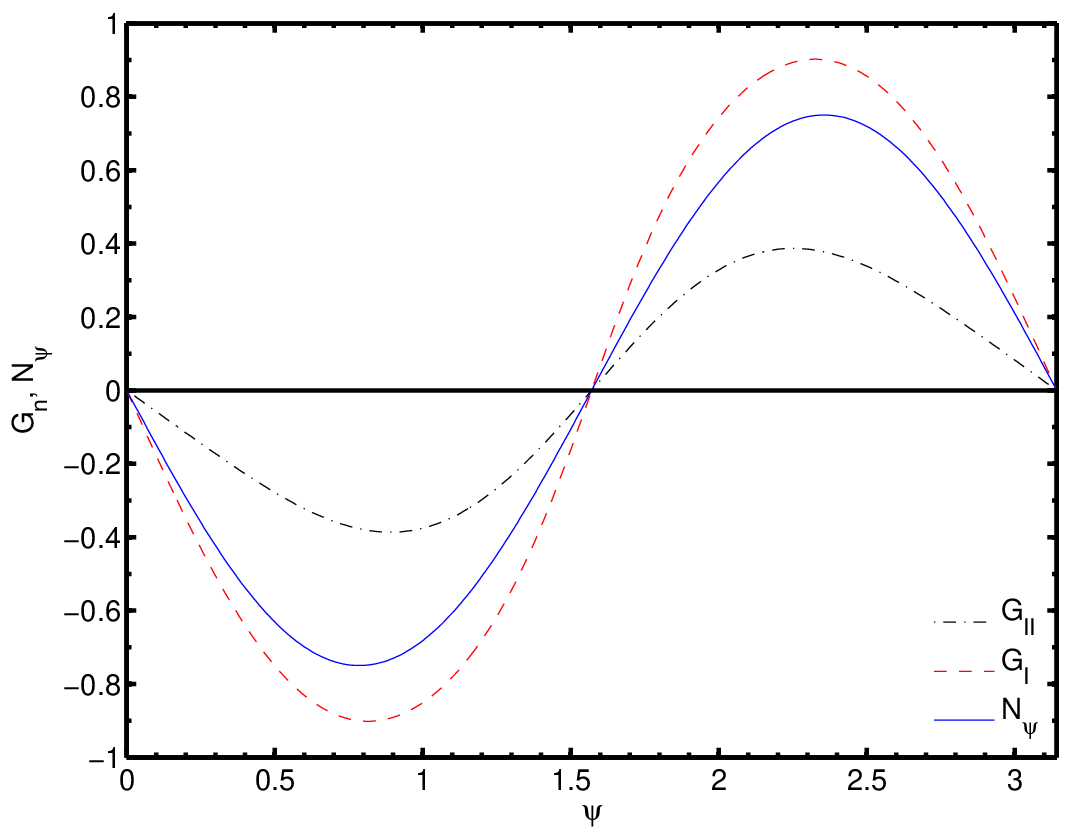}}
\caption{\label{fig:typeIIproifles}{The angular profile functions F and G for the inner bubble forces in both type I and type IIB theories, alongside the linear Newtonian radial and azimuthal angular profiles. }}\end{center} \end{figure} 
and we compare angular profile functions for type I and IIB solutions in Figure \ref{fig:typeIIproifles}.  We see from the form of the MONDian force, \be \nonumber
\delta \vF = -\Del\phi = \frac{4 \pi a_0}{\kappa} \left(\frac{r}{r_0}\right)^p
\vect{S}(\psi) \Rightarrow S_{ij} \propto r^{p-1}\ee where in type I, $p \simeq 0.764$ and in type IIB, $p = 0.5$ - clearly the tidal stresses will
have a sharper divergence as we approach the SP.  This is due to the modified Poisson equation being linear in $\phi$ and so with no curl forces present, the inner bubble solutions are not softened as they are in a non-linear theory.

In addition, at the saddle we have region where $|\vect{g}_N| = 0$ and so
we need to consider solutions to the Laplace equation \be \Del^2 \phi_L = 0\ee which subject to smoothness and continuity conditions being satisfied and regularity at the origin, can be written in general by \be \phi_L = \frac{4 \pi a_0}{\kappa}\sum_\ell A_\ell \, r^\ell\, P_{2\ell}(\cos \psi)\ee where $P_{2\ell}(\cos \psi)$ are Legendre polynomials, \be A_\ell = \frac{a_\ell}{r_0^{\ell-1}}\ee and $a_\ell$ are dimensionless constants to be found by matching solutions at the intermediate MONDian regime (akin to the DM scaling $C$ in type I theories). Our normalisation is picked such that $\Del\phi$ has units of acceleration. We only need to expand out a few terms from this contribution, since the region of validity of these solutions is small.

\subsubsection{QN Regime}
For $r \gg r_0$, we find (\ref{sourcePoisson}) reduces to: %\noindent Here $v > 1$, hence $\nu \approx 1 - \mathcal{O}(1/v^2) + ...$
\be \Del^2\phi = \frac{4\pi a_0}{\kappa}\frac{r_0}{2 r^2 N^2} \left(N_r + \frac{\partial N}{\partial \psi}\frac{N_\psi}{N}\right) \left\{1 - \frac{3}{4} \frac{1}{w^2} + \dots \right\} \label{near QN}\ee where at leading order, we label the source term \be \frac{2 (7 + 9\cos 2\psi)}{(5 + 3\cos 2\psi)^{2}} = h(\psi)\ee In order to satisfy our boundaries conditions, our ansatz for
the leading term needs to be of the form \be \phi_2 = C_1 \,H_2(\psi) + C_2 \ln\left(\frac{r}{r_0}\right) \ee Computing the Laplacian gives \be \Del^2\phi = \frac{C_1}{r^2}\frac{1}{\sin \psi}\frac{\partial }{\partial \psi}(\sin H_2') + \frac{C_2}{r^2}r_0 = \frac{4\pi a_0\, r_0}{\kappa}\frac{h(\psi)}{r^2}  \ee allowing us to set \be C_1 = C_2\, r_0 = \frac{4\pi a_0\,r_0}{\kappa}\ee Integrating out once then gives  \be \sin \psi \,\frac{\partial H_2}{\partial
\psi} = \int (h - 1) \sin \psi\, d\psi + A \label{H2 eqn}\ee and from the
boundary conditions, we find \be A = -\left(\frac{3}{2} + \frac{\pi}{3 \sqrt{3}}\right)\ee
meaning we solve (\ref{H2 eqn}) to find \bea H_2\approx -0.2292 + 0.2876 \cos 2\psi - 0.1163 \cos 4\psi \eea Expanding to higher terms
 will result in the series \be \phi = \phi_2 + \frac{4\pi}{k} a_0 \sum_{n = 2}^{\infty} C_n\left(\frac{r}{r_0}\right)^{2-2n}H_n(\psi) + \frac{\kappa}{4\pi}\Phi_N\ee where $H_n(\psi)$ satisfies the sourced ODE \be n(n+1)H_n + \cot(\psi) H'_n + H''_n = h_n \ee and $h_n$ is given by \be h_n (\psi) = \frac{2^{3n/2-2}(7 + 9\cos 2\psi)}{(5 + 3\cos 2 \psi)^{n/2 + 1}}\ee We also always have the background rescaled Newtonian contribution \be \frac{\kappa}{4\pi}\Phi_{N} = -\frac{\kappa}{4\pi}\frac{A r^2 Nr}{2} = -\frac{4\pi a_0}{\kappa}\frac{r^2}{8 r_0}(1+3\cos 2\psi)\ee which obviously is the dominant contribution in the $r/r_0 \gg 1$ limit.\newline

\section{Adaptations to the Numerical Code}\label{code}
%===============================================================================
The main dynamics of the relaxation code used are outlined in Appendix A
of~\cite{bevis} and here we detail how it can be adapted to solve the type
II Poisson equation.  First we must frame (\ref{type II Laplacian}) in the form of \bea \Del \cdot \vect{g} &=& \frac{k}{4\pi}\Del \cdot\left(\nu\,\vect{g}_N\right) \label{div
g}\\ \Del \wedge \vect{g} &=& 0 \label{curl g}\eea next compute the discrete divergence \bea D_{\vect{x}} = \sum_j \frac{g^j_{\vect{x}} - g^j_{\vect{x-j}}}{\Delta^j_{-}}\eea where we use the compact notation \bea \Delta^j_{-} = r^j_{\vect{x}} - r^j_{\vect{x-j}} \nonumber\\ \Delta^j_{+} = r^j_\vect{x+j} - r^j_\vect{x} \nonumber\eea Finally
the source term takes the form \be D_{\vect{x}}^N = \sum_j \frac{\nu_{\vect{x}}(g^j_{\vect{x}})_N - \nu_{\vect{x-j}}(g^j_{\vect{x-j}})_N}{\Delta^j_{-}} \ee such that at each site we locally solve (\ref{div g}) whilst ensuring the discrete curl (\ref{curl g}) is satisfied globally.  As the field changes $g^j \rightarrow g^j + \delta g^j$, then at each step, these changes should take the form \bea \delta g^j_\vect{x} = +\frac{C_\vect{x}}{\Delta^j_{+}} \nonumber\\ \delta g^j_\vect{x-j} = -\frac{C_\vect{x}}{\Delta^j_{-}} \eea to keep the discrete curl satisfied.  The change in the discrete divergence, $\delta D_\vect{x}$, at each step will satisfy, \be D_{\vect{x}} + \delta D_{\vect{x}} = D_{\vect{x}}^N \ee such at each site, the change
in the field is given by \be \delta g^j_{\vect{x}} = - \frac{D_\vect{x} - D_\vect{x}^N}{\delta D_\vect{x}}\frac{C_\vect{x}}{\Delta^j_{+}}\ee
Then as we cycle through the lattice and the $\vect{g}$ field converges, the additional changes to $g^j_\vect{x}$ lessen.  We achieve faster convergence using a successive over relaxation method (SOR), by scaling the
field as \be \delta g^j_\vect{x} \rightarrow \lambda \delta g^j_\vect{x}\ee
where $\lambda$ is the over-relaxation parameter and is larger than unity. We begin with $\lambda=1$ and increase it once the field is settling down, since high values of $\lambda$ can initially result in the RMS value of $|\delta
D_\vect{x}|$ increasing, whilst we are looking for $|\delta D_\vect{x}| \rightarrow 0$.

%==================================================================================
\bibliography{references}

\end{document}